\documentclass[aps,pre,twocolumn,floats,showpacs]{revtex4}
\usepackage{epsfig}
\usepackage{color}
\usepackage{bm}
\usepackage{latexsym}

\begin{document}
\newcommand{\hide}[1]{}
\newcommand{\tbox}[1]{\mbox{\tiny #1}}
\newcommand{\half}{\mbox{\small $\frac{1}{2}$}}
\newcommand{\sinc}{\mbox{sinc}}
\newcommand{\const}{\mbox{const}}
\newcommand{\trc}{\mbox{trace}}
\newcommand{\intt}{\int\!\!\!\!\int }
\newcommand{\ointt}{\int\!\!\!\!\int\!\!\!\!\!\circ\ }
\newcommand{\eexp}{\mbox{e}^}
\newcommand{\bra}{\left\langle}
\newcommand{\ket}{\right\rangle}
\newcommand{\EPS} {\mbox{\LARGE $\epsilon$}}
\newcommand{\ar}{\mathsf r}
\newcommand{\im}{\mbox{Im}}
\newcommand{\re}{\mbox{Re}}
\newcommand{\bmsf}[1]{\bm{\mathsf{#1}}}
\newcommand{\mpg}[2][1.0\hsize]{\begin{minipage}[b]{#1}{#2}\end{minipage}}

\title{The L\'evy Map: A two-dimensional nonlinear map characterized by
tunable L\'evy flights}

\author{J. A. M\'endez-Berm\'udez,$^1$ Juliano A. de Oliveira,$^2$ and Edson
D. Leonel$^3$}
\affiliation{
$^1$Instituto de F\'{\i}sica, Benem\'erita Universidad Aut\'onoma de Puebla,
Apartado Postal J-48, Puebla 72570, Mexico \\
$^2$UNESP - Univ Estadual Paulista, Campus S\~ao Jo\~ao da Boa Vista,
S\~ao Jo\~ao da Boa Vista, SP 13874-149, Brazil \\
$^3$Departamento de F\'isica, UNESP - Univ Estadual Paulista, Av. 24A, 1515,
Bela Vista, 13506-900 Rio Claro, SP, Brazil
}

\begin{abstract}
Once recognizing that point particles moving inside the extended version of
the rippled billiard perform L\'evy flights characterized by a L\'evy-type
distribution $P(\ell)\sim \ell^{-(1+\alpha)}$ with $\alpha=1$, we derive a
generalized two-dimensional non-linear map $M_\alpha$ able to produce L\'evy
flights described by $P(\ell)$ with $0<\alpha<2$. Due to this property, we
name $M_\alpha$ as the \emph{L\'evy Map}. Then, by applying Chirikov's
overlapping resonance criteria we are able to identify the onset of global
chaos as a function of the parameters of the map. With this, we state the
conditions under which the L\'evy Map could be used as a L\'evy pseudo-random
number generator and, furthermore, confirm its applicability by computing
scattering properties of disordered wires.
\end{abstract}
\maketitle

\section{Introduction and motivation}

The main feature of a L\'evy-type density distribution $P(\ell)$ is the slow,
power-law, decay of its tail. More precisely, for large $\ell$,
\begin{equation}
P(\ell) \sim \frac{1}{\ell^{1+\alpha}} \ ,
\label{levy}
\end{equation}
with $0 < \alpha <2$. Note that the second moment of $P(\ell)$ diverges for
all $\alpha$ and if $0< \alpha <1$ also the first moment diverges. This kind
of distributions are also known as $\alpha$-stable distributions
\cite{uchaikin}. Random processes characterized by probability densities with
a long tail (L\'evy-type processes) have been found and studied in very
different phenomena and fields such as biology, economy, and physics. Among
many of recently studied systems showing L\'evy-type processes we can mention:
animal foraging \cite{VRL08}, human mobility \cite{BHG06}, earthquake
statistics \cite{SAP00}, mosquitoes flight under epidemiological modeling
\cite{botari}, and light transmission through a disordered glass
\cite{BBW08}. See also \cite{CSN11} for a compilation of systems displaying
L\'evy flights.

In particular, to help us to introduce later the main model system of this study, 
i.e. the \emph{L\'evy Map}, we want to describe in some detail a simple dynamical 
model characterized by L\'evy processes: the \emph{ripple billiard}.

\begin{figure}[b]
\centerline{\includegraphics[width=7cm]{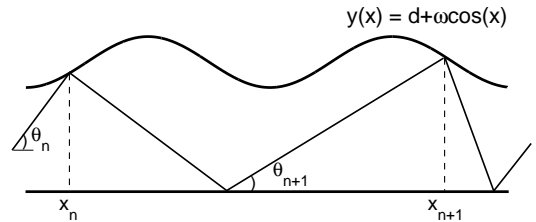}}
\caption{Geometry of the rippled billiard and definition of the variables of map $M$, see
Eq.~(\ref{M}).}
\label{Fig1}
\end{figure}

\begin{figure*}[t]
\centerline{\includegraphics[width=11cm]{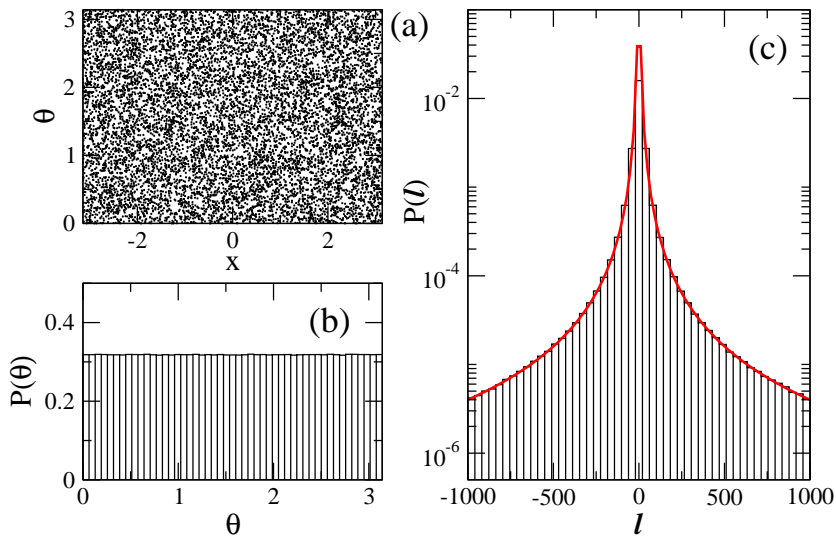}}
\caption{(Color online) (a) Poincar\'e map $M$, (b) phase distribution $P(\theta)$, and 
(c) length distribution $P(\ell)$ for the ripple billiard with $\omega=2\pi/10$ and $d=2\pi$. 
A single initial condition $x_0=\theta_0=0.1$ was iterated (a) $10^4$ and (b-c) $10^7$ times. 
The red full curve in (c) is Eq.~(\ref{levyM}).}
\label{Fig2}
\end{figure*}

The ripple billiard, see for example Chapter 6 of \cite{linda}, consists of two walls: 
one flat at $y=0$ and a rippled one given by the function $y=d+\omega\cos(x)$; 
here $d$ is the average width of the billiard and $\omega$ the ripple amplitude, see Fig.~\ref{Fig1}. 
An attractive feature
of the ripple billiard is that its classical phase space undergoes the generic transition to
global chaos as the amplitude of the cosine function increases. Then, results from the analysis of 
this system are applicable to a large class of systems, namely non-degenerate, non-integrable 
Hamiltonians \cite{linda,licht}. Moreover, the dynamics of
classical particles inside the ripple billiard can be well approximated by a two-dimensional (2D)
Poincar\'e map $M$ between successive collisions with the rippled boundary \cite{german,edson}
$(\theta_{n+1}, x_{n+1}) = M(\theta_n, x_n)$, where $\theta_n$ is the angle the particle's
trajectory makes with the $x$-axis just before the $n$th bounce with the rippled boundary
at $x_n$. Map $M$ can be easily derived and, after the assumptions $\omega\sin(x_n)\ll 1$ and $\omega/d\ll1$,
it gets the simple form
\begin{equation}
M:\left\{\begin{array}{ll}
\theta_{n+1} & = \theta_n - 2\omega\sin(x_n) \ , \nonumber \\
x_{n+1} & = x_n + 2d\cot(\theta_{n+1}) \ . \
\end{array}
\right.
\label{M}
\end{equation}
As an example, in
Fig.~\ref{Fig2}(a) we plot the Poincar\'e map $M$ for $\omega=2\pi/10$ and $d=2\pi$. It is clear from
this plot that this combination of geometrical parameters produces ergodic dynamics (also known as 
global chaos). Notice that
in this figure we have plotted the variable $x$ modulus $2\pi$, as usual for this kind of 2D
maps; with this, we can globally visualize the map dynamics in a single plot but we may loose
important information.


Among the dynamical information which is lost when applying $\mbox{mod}(2\pi)$ to a map such as
$M$, we can mention the length of paths between successive map iterations
$\ell\equiv x_{n+1}-x_n$, i.e. the length between two successive collisions with the rippled
boundary of the billiard. In fact, in Fig.~\ref{Fig2}(c) we present $P(\ell)$ for the same
parameters used to construct Fig.~\ref{Fig2}(a). From this figure we can
clearly see that: (i) even though most of the paths $\ell$ produced by map $M$
are short (i.e. $P(\ell)$ is highly
peaked at $P(\ell)=0$), there is a non-negligible probability for very large values of $\ell$ to
occur: Notice that the values of $\ell=\pm 1000$ at the edges of Fig.~\ref{Fig2}(c) mean that a
particle has traveled about 160 periods of the rippled billiard between two successive collisions
with the rippled boundary and; (ii) $P(\ell)$ decays as a power-law similar to
Eq.~(\ref{levy}).
These two facts are explicit evidences of L\'evy flights in the dynamics of map $M$.
Thus, the following question becomes pertinent: Can we provide an analytic expression for the
shape of $P(\ell)$ given the simple form of map $M$? Fortunately, the answer is positive as we
will show it below.

If we consider the dynamics of map $M$ to be in the regime of full chaos then a single
trajectory can explore the full available phase space homogeneously,
as shown in Fig.~\ref{Fig2}(a), so $P(\theta)$ is constant and equal to $1/\pi$,
as verified in Fig.~\ref{Fig2}(b).
Also, from the second equation in map $M$ we obtain $\theta=\tan^{-1}(2d/\ell)$.
Thus, using $P(\ell) = P(\theta) | d\theta/d\ell |$, we can write
\begin{equation}
P(\ell) = \frac{2d}{\pi\left( 1 + \ell^2 \right)} \ ,
\label{levyM}
\end{equation}
which is in fact a L\'evy-type probability distribution function with $\alpha=1$; compare with
Eq.~(\ref{levy}). Then, in Fig.~\ref{Fig2}(c) we plot Eq.~(\ref{levyM}) (as the red full line)
together with the numerically obtained $P(\ell)$ and observe a very good correspondence making
clear the existence of L\'evy processes, characterized by the power-law decay $\alpha=1$, in the
dynamics of the rippled billiard.

In fact, the origin of the L\'evy-type probability distribution of Eq.~(\ref{levyM}) for the 
lengths $\ell$ in the ripple billiard is the existence of L\'evy flights.
Since a typical chaotic 
trajectory fills the available phase space uniformly, see Fig.~\ref{Fig2}(a), then all angles
$\theta\in(0,2\pi)$ are equally probable; however, different angles produce quite different 
lengths $\ell$. For example, an angle $\theta$ very close to $\pi/2$ corresponds to 
a very short length $\ell\sim 0$, see Fig.~\ref{Fig1}. While $\theta$ tending to zero or $\pi$ 
produce trajectories which are nearly parallel to the $x$-axis that may travel very long distances 
between successive collisions with the ripple boundary: in such case $\ell\to\infty$. 
These {\it grazing} trajectories are indeed L\'evy flights, known to produce heavy-tailed 
distribution functions \cite{M82}; for the ripple billiard the L\'evy flights produce 
Eq.~(\ref{levyM}) as derived above.
Moreover, grazing trajectories in the ripple billiard have been found to play a prominent 
role when defining the classical analogs of the quantum {\it structure of eigenstates} and 
{\it local density of states} \cite{LMI00}.

Equation~(\ref{levyM}) is already an interesting result on the dynamics of the rippled billiard 
(and of general chaotic extended billiards with infinite horizon) that deserves additional attention,
however our goal here is different. Since now we know that map $M$ produces L\'evy flights
characterized by $\alpha=1$ we ask ourselves: Can we propose a general 2D non-linear map where
$\alpha$ can be included as a parameter? More generally, can we construct the map $M_\alpha$ able
to produce L\'evy flights characterized by $0<\alpha<2$? Indeed, in the following section we
elaborate on these questions.

\section{Derivation of the L\'evy Map}

We introduce the \emph{L\'evy Map} $M_\alpha$ by following the opposite procedure 
we used above to obtain the distribution function of Eq.~(\ref{levyM}) from map $M$.

Let us
\begin{itemize}
  \item[(i)]
consider the 2D map $(\theta_{n+1}, x_{n+1}) = M_\alpha(\theta_n, x_n)$ to have the 
same iteration relation for the angle $\theta$ as map $M$, see Eq.~(\ref{M}),
  \item[(ii)]
assume the map $M_\alpha$ to be in a regime of global chaos, such that 
\begin{equation}
P(\theta)=\mbox{const}=\frac{1}{\pi} \ ,
\label{Pofteta}
\end{equation}
and
  \item[(iii)]
demand the variable 
\begin{equation}
\ell\equiv x_{n+1}-x_n
\label{ell}
\end{equation} 
from map $M_\alpha$ to be characterized by the L\'evy-type density distribution function
\begin{equation}
P(\ell) = \frac{\cal C}{\ell^{1+\alpha}} \ ,
\label{levy2}
\end{equation}
where $0<\alpha<2$ and $\cal C$ is a normalization constant.
\end{itemize}
Then,
\[
\theta \equiv \int \frac{P(\ell)}{P(\theta)} d\ell = 
\pi{\cal C} \int \frac{d\ell}{\ell^{1+\alpha}} = 
-\frac{\pi{\cal C}}{\alpha\ell^\alpha}
\]
provides $\ell=(-\alpha\theta/\pi{\cal C})^{-1/\alpha}$. Therefore, we define the 
\emph{L\'evy Map} as
\begin{equation}
M_\alpha:\left\{\begin{array}{ll}
\theta_{n+1} & = \theta_n - 2\omega \sin(x_n) \ , \nonumber \\
x_{n+1} & = x_n + \gamma|\alpha\theta_{n+1}|^{-1/\alpha} \ , \quad 0 < \alpha <2 \ ,
\end{array}
\right.
\label{Malpha}
\end{equation}
where $\omega$, $\gamma=(\pi{\cal C})^{1/\alpha}$, and $\alpha$ are the map parameters. 
We have 
introduced the absolute value in the second equation of $M_\alpha$ to avoid fractional 
powers of negative angles. This, in turn, makes all {\it lengths} $\ell$ to be positive.

Notice that for $\alpha=1$ and $\theta_{n+1}\ll 1$, where 
$\cot(\theta_{n+1})\approx 1/\theta_{n+1}$, we recover map $M$ from $M_\alpha$ (with 
$\gamma=2d$). We also note that $M_\alpha$ has a similar form than the maps studied in 
Refs.~\cite{OBL10,LOS11,OL11,ODCL13} in the sense that the function $f(\theta_{n+1})$, 
in the second line of the map $M_\alpha$, is inverse proportional to $\theta_{n+1}$ to 
a non-integer power. 

Below we will focus our attention on map $M_\alpha$ with the parameter $\alpha$ into
the interval $0<\alpha<2$ because our motivation is to construct a map able to produce 
pseudo-random variables distributed according to $\alpha$-stable distributions. 
However, the parameter $\alpha$ may also take values outside this interval.

\section{Onset of global chaos}

In general, depending on the values of the parameters $(\omega,\gamma,\alpha)$, the 
dynamics of the L\'evy Map may be integrable, mixed (where the phase space contains 
periodic islands surrounded by chaotic seas which may be limited by invariant spanning 
curves), or ergodic. That is, the classical phase space of map $M_\alpha$ develops 
the generic transition to global chaos (not shown here). However, for Eq.~(\ref{Pofteta}) 
to be valid the map dynamics must be ergodic. Therefore, below, by applying Chirikov's 
overlapping resonance criteria we shall identify the onset of global chaos as a function 
of the parameters of the L\'evy Map.

Following \cite{licht} we linearize $M_\alpha$ around the period-one fixed points 
$(x^*,\theta^*)$, which are defined through
\[
\left\{
\begin{array}{ll}
\theta_{n+1} = \theta_n = \theta^* \\
x_{n+1} = x_n = x^* + 2\pi m \ , \quad m = 1, 2, 3, \ldots \\
\end{array}
\right.
\]
This condition provides 
\begin{equation}
x^*=0,\pi \qquad \mbox{and} \qquad 
\theta^* = \frac{1}{\alpha}\left(\frac{\gamma}{2\pi m}\right)^\alpha \ .
\label{FP}
\end{equation} 
Then, for an angle close to $\theta^*$ we can write $\theta_n = \theta^*+\Delta \theta_n$ 
getting 
\begin{eqnarray}
\theta_{n+1} & = & \theta^*+\Delta \theta_{n+1} \nonumber \\
& = & \theta^* + \Delta\theta_n - 2\omega\sin(x_n) \nonumber \ .
\end{eqnarray}
Thus, 
\begin{equation}
\Delta \theta_{n+1} = \Delta\theta_n - 2\omega\sin(x_n) \ .
\label{deltateta}
\end{equation}
In addition, for $x$ we have 
\begin{eqnarray}
x_{n+1} & = & x_n + \gamma\alpha^{-1/\alpha}(\theta^*+\Delta\theta_{n+1})^{-1/\alpha} \nonumber \\
& \approx & x_n + \gamma(\alpha\theta^*)^{-1/\alpha}[1-(\alpha\theta^*)^{-1}\Delta\theta_{n+1}] \nonumber \\
& = & x_n + 2\pi m [1-(\gamma/2\pi m)^{-\alpha}\Delta\theta_{n+1}] \nonumber \\
& = & x_n - \gamma^{-\alpha} (2\pi m)^{\alpha+1}\Delta\theta_{n+1} \ .
\label{xn1}
\end{eqnarray}
Finally, by substituting the new angle 
\[
\Theta_n \equiv -\gamma^{-\alpha}(2\pi m)^{\alpha+1}\Delta\theta_n
\]
in (\ref{deltateta}) and (\ref{xn1}) we can write the linearized map
\begin{equation}
M_{\tbox{SM}}:\left\{\begin{array}{ll}
\Theta_{n+1} = \Theta_n + K \sin(x_n) \\
x_{n+1} = x_n + \Theta_{n+1} \\
\end{array}
\right. \ ,
\label{MSM}
\end{equation}
where $\Theta$ and $x$, respectively, play the role of action-angle variables in the 
{\it Standard Map} \cite{licht,C69} with 
\begin{equation}
K = 2(2\pi m)^{\alpha+1}\omega\gamma^{-\alpha} \ , \quad m = 1, 2, 3, \ldots
\label{K}
\end{equation} 
being the stochasticity parameter. 

Chirikov's overlapping resonance criteria predicts the transition to global chaos 
for $K>K_{\tbox{C}}$, where $K_{\tbox{C}}\approx 0.971635\ldots$ 
\cite{licht,C69,C79}. Global chaos means that chaotic regions are
interconnected over the whole phase space (stability islands may still exist
but are
sufficiently small that the chaotic sea extends throughout the vast majority of phase 
space). This criteria for the L\'evy Map reads as 
\begin{equation}
\omega \stackrel{>}{\sim} \frac{\gamma^{\alpha}}{2(2\pi)^{\alpha+1}} = \frac{{\cal C}}{4(2\pi)^\alpha} \ .
\label{caosg}
\end{equation}
In fact, to get Eq.~(\ref{caosg}) from Eq.~(\ref{K}) we have applied the resonance 
criteria to the period-one fixed point corresponding to $m=1$, see Eq.~(\ref{FP}), 
which is the fixed point having the largest $\theta$ (i.e. it is located highest 
in phase space) and the one closer to the last invariant spanning curve bounding 
the diffusion of the variable $\theta$.

\begin{figure}[t]
\centerline{\includegraphics[width=8cm]{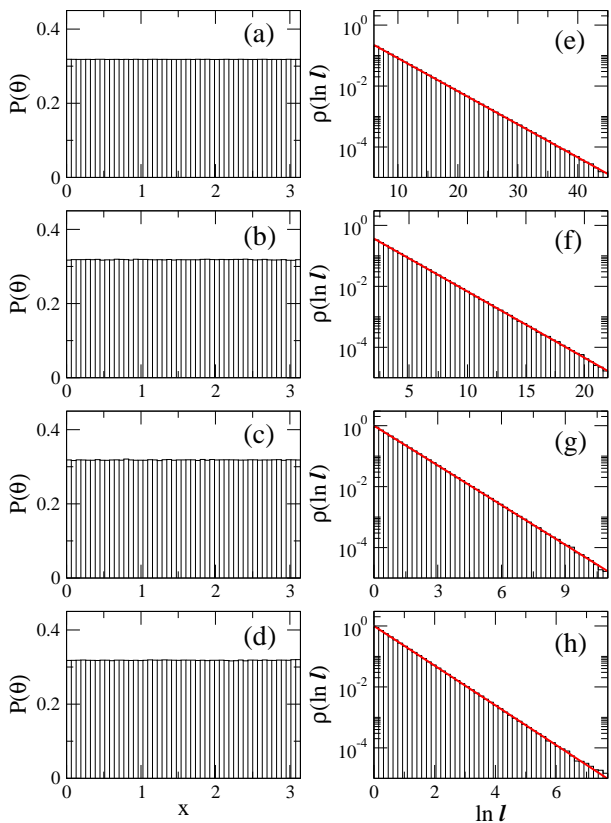}}
\caption{(Color online) (a-d) Phase distributions $P(\theta)$ and (e-h) length 
distributions $\rho(\ln\ell)$ for the L\'evy Map $M_\alpha$ with $\omega={\cal C}=1$. 
(a,e) $\alpha=1/4$, (b,f) $\alpha=1/2$, (c,g) $\alpha=1$, and 
(d,h) $\alpha=3/2$. To construct each histogram a single initial condition 
$x_0=\theta_0=0.1$ was iterated $10^7$ times. 
The red full curve in (e-h) is $\rho(\ln\ell)=\ell^{-\alpha}$.}
\label{Fig4}
\end{figure}

Indeed, we have verified that the phase space of $M_\alpha$ is ergodic if condition 
(\ref{caosg}) is satisfied (not shown here). Moreover, in Figs.~\ref{Fig4}(a-d) we 
plot the phase distribution functions $P(\theta)$ for the L\'evy Map with $\omega={\cal C}=1$ 
corresponding to $\alpha=1/4$, 
1/2, 1, and 3/2. From these figures, it is clear that $P(\theta)$ is certainly a constant 
distribution. In particular, note that with $\omega={\cal C}=1$ condition (\ref{caosg}) is 
satisfied for any $\alpha$, so we shall use this set of parameter values in all figures below.

\begin{figure}[t]
\centerline{\includegraphics[width=7cm]{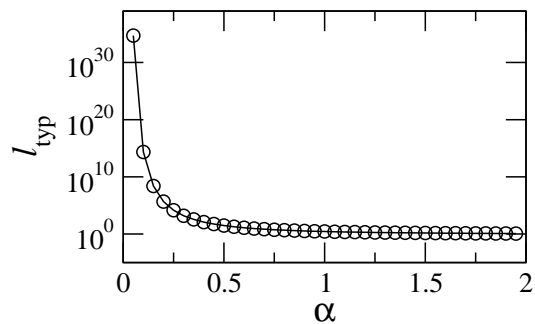}}
\caption{Typical value of $\ell$, $\ell_{\tbox{typ}}=\exp\bra\ln\ell\ket$, 
for the L\'evy Map $M_\alpha$ as a 
function of $\alpha$. $\omega={\cal C}=1$ were used. The average was
performed over $10^7$ values of $\ell$ obtained by iterating $M_\alpha$ from the single 
initial condition $x_0=\theta_0=0.1$.}
\label{Fig5}
\end{figure}

Now we would like to verify that once $P(\theta)=1/\pi$, $M_\alpha$ must produce lengths 
$\{\ell\}$ distributed according to Eq.~(\ref{levy2}). However, we noticed that for 
$\alpha<1$ the L\'evy Map produces huge values of $\ell$. To show this, in Fig.~\ref{Fig5}
we plot the typical value of $\ell$, 
\begin{equation}
\ell_{\tbox{typ}} = \exp\bra\ln\ell\ket \ ,
\label{elltyp}
\end{equation}
as a function of $\alpha$; where we can observe that for $\alpha=1/4$ the typical $\ell$ is
larger than $10^5$ (in fact, from the data we used to construct the $P(\theta)$ of 
Fig.~\ref{Fig4}(a) we obtained several lengths $\ell$ of the order of
$10^{30}$!). Thus, it is
not practical to construct $P(\ell)$ to test the validity of Eq.~(\ref{levy2}) itself.
Instead, we make the change of variable $\ell\to\ln\ell$ which leads to 
\[
\rho(\ln\ell)=\ell P(\ell)=\frac{{\cal C}}{\ell^{\alpha}} \ .
\]
Then in Figs.~\ref{Fig4}(e-h) we show length distribution functions $\rho(\ln\ell)$ for 
the L\'evy Map with $\alpha=1/4$, 1/2, 1, and 3/2 (histograms). As clearly seen, the 
agreement between the histograms and $\rho(\ln\ell)=\ell^{-\alpha}$ (shown as red 
thick lines) is indeed excellent.

It is relevant to stress that since the phase space of the L\'evy Map is ergodic when 
condition (\ref{caosg}) is satisfied, the sequences $\{\ell\}$ can then be considered 
as L\'evy-distributed pseudo-random numbers. In fact, in the next Section we will show 
through a specific application that the lengths $\ell$ can be used in practice as 
random numbers.

\section{The L\'evy Map as a L\'evy-distributed pseudo-random number generator}
\label{conductance}

There is a good deal of work devoted to the use of non-linear maps as pseudo-random 
number generators, see some examples in Refs.~\cite{pseudo1,pseudo2,pseudo3,pseudo4,pseudo5,pseudo6}.
Therefore, in a similar way, we would like to use the L\'evy Map to generate pseudo-random numbers 
particularly distributed according to the L\'evy-type probability distribution function 
of Eq.~(\ref{levy2}). However, instead of analyzing the randomness of the sequences $\{\ell\}$ 
produced by $M_\alpha$, here we will show that these numbers can be successfully used already in 
a specific application: We shall compute transmission through one-dimensional (1D) disordered wires.

Recently, the electron transport through 1D quantum wires with L\'evy-type 
disorder was studied in Refs.~\cite{FG10,FG12}. There, it was found that the average (over 
different disorder realizations) of the logarithm of the dimensionless conductance $G$ 
behaves as
\begin{eqnarray}
\left\langle  - \ln G \right\rangle \propto \left\{\begin{array}{ll}
L^\alpha & \mbox{for} \quad 0 < \alpha < 1  \\
L        & \mbox{for} \quad 1 \le \alpha < 2
\end{array}\right. \ .
\label{G}
\end{eqnarray}
where $L$ is a {\it length} that depends on the wire model. For example, for a wire represented
as a sequence of potential barriers with random lengths, $L=\sum_n \nu_n$ \cite{FG10}; where 
$\nu_n$ is the length of the $n$th barrier in the wave propagation direction. While for a wire 
represented by the 1D Anderson model with off-diagonal disorder, $L=\sum_n\nu_{n,n+1}$ \cite{FG12}; 
where $\nu_{n,n+1}$ is the hopping integral between the sites $n$ and $n+1$.

\begin{figure}[t]
\centerline{\includegraphics[width=6cm]{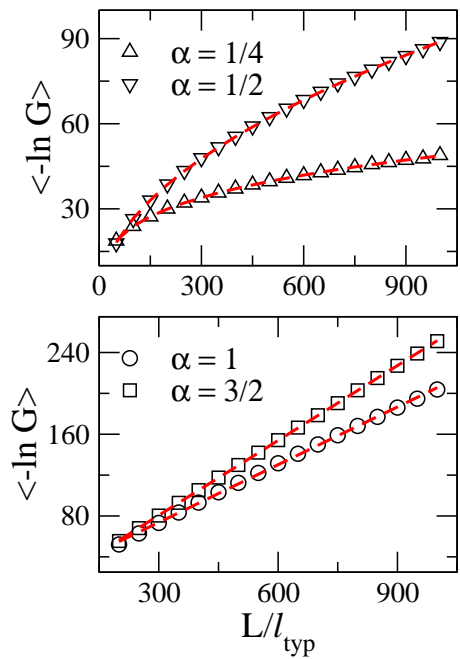}}
\caption{The average $\bra-\ln G\ket$ as a function of $L/\ell_{\tbox{typ}}$
for the 1D Anderson Model with off-diagonal L\'evy-type disorder characterized
by $\alpha$. We used an incoming wave with energy $E=0.1$. The dashed lines are 
fittings of the data with Eq.~(\ref{G}). Each symbol was calculated using $10^5$ 
wire realizations. Each wire realization is constructed from a single sequence of 
lengths $\{\ell\}$ generated by map $M_\alpha$ having random initial conditions 
uniformly distributed in the intervals $-\pi<x_0<\pi$ and $0<\theta_0<2\pi$.}
\label{Fig6}
\end{figure}

Here, we use the 1D Anderson model with off-diagonal disorder to represent 1D quantum wires 
(see details in the Appendix) where the hopping integrals $\nu_{n,n+1}$ are, in fact, 
the pseudo-random lengths $\ell$ generated by our L\'evy Map. Then, in Fig.~\ref{Fig6} we plot 
$\bra-\ln G\ket$ as a function of $L$ for the 1D Anderson Model with L\'evy-type disorder 
characterized by $\alpha=1/4$, 1/2, 1, and 3/2. We have computed the dimensionless conductance 
by the use of the effective Hamiltonian approach (see details in the Appendix). 
In Fig.~\ref{Fig6} we are using $\ell_{\tbox{typ}}$ to normalize $L$ to be able to show curves 
corresponding to different values of $\alpha$ into the same figure panel. Also, in Fig.~\ref{Fig6} 
we are including fittings of the curves $\bra-\ln G\ket$ vs.~$L$ with Eq.~(\ref{G}), see red dashed 
lines, which certainly show the {\it anomalous} conductance behavior predicted in 
Refs.~\cite{FG10,FG12}. Therefore, validating in this way the use of the L\'evy Map as a 
pseudo-random number generator.
 
\section{Conclusions}

In this paper we have introduced the so-called {\it L\'evy Map}: A
two-dimensional nonlinear map characterized by tunable L\'evy flights. Indeed
it is described by a 2D nonlinear and area preserving map with a control
parameter driving two important transitions: (i) integrability
($\omega=0$) to non-integrability ($\omega\ne 0$) and; (ii) local chaos with
$\omega<C/4(2\pi)^{\alpha}$ to globally chaotic dynamics with
$\omega>C/4(2\pi)^{\alpha}$. We have applied Chirikov's overlapping
resonance criteria to identify the onset of global chaos as a function of the
parameters of the map therefore reaching to condition (ii) as described on
the line above. In this way we stated the requirements under which the L\'evy
Map could be used as a L\'evy pseudo-random number generator and confirmed
its effectiveness by computing scattering properties of disordered wires.

\section{Appendix}

In Sect.~\ref{conductance} we have considered 1D tight-binding chains of size $N$ 
described by the Hamiltonian
\begin{equation}
 H_{mn} = h_n\delta_{mn} + \nu_{n,n+1}\delta_{n,n+1}+ \nu_{n,n-1}\delta_{n,n-1} \ ,
 \label{Hmn}
\end{equation}
where $h_n$ are on-site potentials that we set to zero and $\nu_{n,n+1}=\nu_{n+1,n}$ 
are hopping amplitudes connecting nearest sites. Here, $m,n=1\ldots N$.

We open the 1D chains by attaching two single-mode semi-infinite leads to the opposite
sites on the 1D samples. Each lead is described by the 1D semi-infinite 
tight-binding Hamiltonian
\[
H_{\tbox{lead}}=\sum^{-\infty}_{n=1} (| n \rangle \langle n+1| + |n+1 \rangle \langle n|) \ .
\]
Then, following the {\it effective Hamiltonian approach}, the scattering matrix 
($S$-matrix) has the form \cite{MW69} 
\begin{equation}
\label{smatrix}
S =
\left(
\begin{array}{cc}
r & t'   \\
t & r'
\end{array}
\right)
={\bf 1}-2\pi\mbox{i} \, {\cal W}^{\,T} \frac{1}{E-{\cal H}_{\rm eff}} {\cal W} \ .
\end{equation}
where $t$, $t'$, $r$, and $r'$ are transmission and reflection amplitudes; ${\bf 1}$ is 
the $2\times 2$ unit matrix, $k=\arccos(E/2)$ is the wave vector supported in the leads, 
and ${\cal H}_{\rm eff}$ is an effective energy-dependent non-hermitian Hamiltonian given by
\begin{equation}
\label{Heff}
{\cal H}_{\rm eff} = H + \pi\cot(k){\cal W}{\cal W}^{\,T} - \mbox{i}\pi {\cal W}{\cal W}^{\,T} \ .
\end{equation}
Above, ${\cal W}$ is a $L\times 2$ matrix with elements 
${\cal W}_{i,j}=[\sin(k)/\pi]^{1/2}(\delta_{1,1}+\delta_{L,2})$.

Finally within a scattering approach to electronic transport, once the scattering matrix
is known we compute the dimensionless conductance as \cite{Landauer} 
\begin{equation}
G=|S_{12}|^2 \ .
\end{equation}

{\bf Acknowledgments.}
J.A.M.-B is grateful to FAPESP (2013/14655-9) Brazilian agency; partial
support form VIEP-BUAP grant MEBJ-EXC14-I and Fondo Institucional PIFCA 2013
(BUAP-CA-169) is also acknowledged. 
J.A.M.-B also thanks the warm hospitality at Departamento de F\'isica
at UNESP -- Rio Claro, where this work was mostly developed. J.A.O thanks
PROPe/FUNDUNESP/UNESP. E.D.L. thanks to FAPESP (2012/23688-5), CNPq, and CAPES, 
Brazilian agencies.

\end{document}